\documentclass[final,5p,times,twocolumn]{elsarticle}
\usepackage{amsmath}
\usepackage{bm}
\usepackage{booktabs}
\usepackage{braket}
\usepackage{caption}
\usepackage{color}
\usepackage{graphicx}
\usepackage[colorlinks]{hyperref}
\usepackage{multirow}

\journal{Journal of Subatomic Particles and Cosmology}

\begin{document}

\begin{frontmatter}
  \title{A dual description of quarks and baryons: \\ Quarkyonic matter within a relativistic quark model}
  \author[1]{Tsuyoshi Miyatsu}
  \address[1]{Department of Physics and OMEG Institute, Soongsil University, Seoul 06978, Republic of Korea}
  \ead{tsuyoshi.miyatsu@ssu.ac.kr}
  \author[1]{Myung-Ki Cheoun}
  \ead{cheoun@ssu.ac.kr}
  \author[2]{Koichi Saito}
  \address[2]{Department of Physics and Astronomy, Faculty of Science and Technology, Tokyo University of Science, Noda 278-8510, Japan}
  \ead{koichi.saito@rs.tus.ac.jp}

  \begin{abstract}
    We investigate quarkyonic matter within a relativistic quark model by combining the dual quarkyonic picture with the quark-meson coupling (QMC) model.
    Using relativistic gaussian quark wavefunctions for the nucleon, we construct the quarkyonic QMC (QQMC) model and study the properties of symmetric nuclear matter and pure neutron matter.
    We find that the quark saturation density depends sensitively on the nucleon size parameter and that nuclear interactions quantitatively modify the high-density behavior of the equation of state (EoS) and the sound velocity.
    In particular, the QQMC model yields an earlier onset of quark saturation than the noninteracting gaussian quarkyonic (GQ) model, indicating that nuclear interactions enhance the stiffening of the EoS in the quarkyonic regime.
  \end{abstract}


\end{frontmatter}

\section{Introduction}
\label{sec:introduction}

Neutron stars provide a unique opportunity to investigate the properties of strongly interacting matter at densities beyond the nuclear saturation density.
Recent astrophysical observations, including precise measurements of neutron-star masses, radius estimates from the NICER mission, and tidal-deformability constraints from gravitational-wave events, have significantly narrowed the possible behavior of the equation of state (EoS) of dense matter~\citep{Vinciguerra:2023qxq,Choudhury:2024xbk,Salmi:2024aum,LIGOScientific:2017vwq,LIGOScientific:2018hze}.
% ~\citep{Vinciguerra:2023qxq,Choudhury:2024xbk,Salmi:2024aum,Salmi:2024bss,Mauviard:2025dmd,LIGOScientific:2017vwq,LIGOScientific:2018hze}.
In particular, the existence of neutron stars with masses around two solar masses implies that the EoS must remain sufficiently stiff in the high-density region~\citep{NANOGrav:2017wvv,NANOGrav:2019jur}.
% ~\citep{Demorest:2010bx,Antoniadis:2013pzd,NANOGrav:2017wvv,NANOGrav:2019jur}.
For this reason, clarifying the composition and thermodynamic properties of dense matter has become one of the central issues in nuclear astrophysics.

From a theoretical point of view, however, the description of dense matter cannot be limited to hadronic degrees of freedom alone.
As the baryon density increases, the internal quark structure of hadrons is expected to play an important role, and explicit quark degrees of freedom may eventually become unavoidable~\citep{Miyatsu:2011bc,Miyatsu:2013yta,Whittenbury:2013wma}.
Moreover, in the asymptotically high-density limit, dense QCD matter should be connected to the perturbative QCD (pQCD) regime~\citep{Kurkela:2014vha}.
This requirement has motivated many studies of dense-matter EoSs that incorporate quark matter in addition to hadronic matter, with the aim of constructing a consistent description from nuclear density to the quark-dominated regime~\citep{Weber:2004kj}.
% ~\citep{Haensel:1986qb,Weber:2004kj}.

A major unresolved issue in this context is how the transition from hadronic matter to quark matter actually occurs.
Depending on the theoretical framework, the transition may be described as a first-order phase transition~\citep{Glendenning:1992vb}, a smooth crossover~\citep{Masuda:2012kf}, or a quarkyonic-like regime in which baryonic and quark degrees of freedom coexist in a dual manner~\citep{MCLERRAN200783}.
To investigate such possibilities, it is important to use a framework that can simultaneously account for nuclear interactions and the internal quark structure of baryons.
The quark-meson coupling (QMC) model is particularly suitable for this purpose, because it explicitly incorporates the quark substructure of the nucleon through the coupling of confined quarks to meson mean fields, thereby providing a bridge between nuclear many-body dynamics and quark-level effects~\citep{Saito:1994ki,Guichon:1995ue,Saito:2005rv,Guichon:2018uew}.

In the present study, we formulate the nucleon in terms of relativistic quark wavefunctions and use this description as the basis for extending the QMC model to dense matter in a quarkyonic-like regime.
This approach therefore enables us to investigate dense matter by consistently combining relativistic quark dynamics, nuclear interactions, and quarkyonic features.
We focus on symmetric nuclear matter (SNM) and pure neutron matter (PNM), and study how the quark saturation emerges and how the interactions modify the nuclear EoS and the sound velocity at high density.
In particular, by comparing the interacting quarkyonic QMC (QQMC) model with the noninteracting gaussian quarkyonic (GQ) model, we clarify the quantitative role of nuclear interactions in the stiffening of dense matter.

\section{Quark-meson coupling model}
\label{sec:QMC}

We use a gaussian wavefunction for a quark in the nucleon, obtained from the relativistic confinement potential of the scalar-vector harmonic oscillator (HO) type
\begin{equation}
  U(r) = \frac{c}{2} (1+\gamma_0) r^2 ,
  \label{eq:conpot}
\end{equation}
with the strength parameter $c$~\citep{PhysRevD.110.113001}.
The Dirac equation with the HO potential, Eq.~\eqref{eq:conpot}, can be solved analytically.
In free space, the lowest-state solution for an isospin-symmetric quark system ($u=d$) is given by
\begin{align}
  \psi_{Q}({\bm r})
  & = \frac{1}{\pi^{3/4} a^{3/2}} \sqrt{\frac{2 \lambda^2 a^2}{2\lambda^2 a^2+3}}
    \begin{pmatrix}
      1 \\
      i {\vec \sigma} \cdot {\hat r} \frac{1}{\lambda a} \left( \frac{r}{a} \right)
    \end{pmatrix}
  e^{-r^2/2a^2} \chi ,  \label{eq:sol} \\
  \lambda
  & = \epsilon + m, \ \ \  a^2 = \frac{1}{\sqrt{c\lambda}} = \frac{3}{\epsilon^2-m^2} , \label{eq:norm}
\end{align}
where $m$ is the constituent quark mass, $a$ is the length scale and $\chi$ is the quark spinor.  Here, the single-particle quark energy, $\epsilon$, is determined by $\sqrt{\epsilon + m} (\epsilon - m) = 3 \sqrt{c}$.
% Note that, in the limit $\lambda \to \infty$, this wavefunction turns out to be the nonrelativistic (NR) one.
The nucleon mass in vacuum, $M_N$, is thus given by $M_N = E_N^0 + E^{spin}_N - E^{c.m.}_N$ with the zeroth-order energy, $E_N^0 = 3 \epsilon$, and some corrections to the nucleon energy such as the spin correlations, $E^{spin}_N$, due to the quark-gluon and quark-pion interactions, and the center of mass (c.m.) correction, $E^{c.m.}_N$~\cite{PhysRevD.110.113001}.
The parameters and the quark energies are summarized in Table~\ref{tab:parameters}.
In the present relativistic quark model, the nucleon is described only by the valence-quark contribution.
Therefore, the root-mean-square charge radius of proton, $r_{p}$, should not be identified directly with the empirical proton charge radius, and somewhat smaller values are adopted in this study.
\begin{table}[t!]
  \centering
  \caption{\label{tab:parameters}
    Parameters in the relativistic quark model of nucleon in vacuum.
    We present three cases of $r_{p}=0.6$, $0.7$, and $0.8$ fm.
    The constituent quark mass is fixed at $m=300$ MeV.}
  \begin{tabular}{ccccccc}
    \toprule
    $r_{p}$ & $a$ & $c$ & $\epsilon$ & $E_{N}^{0}$ & $E_{N}^{spin}$ & $E_{N}^{c.m.}$ \\
    \cline{1-2}\cline{4-7}
    \multicolumn{2}{c}{(fm)} & (fm$^{-3}$) & \multicolumn{4}{c}{(MeV)} \\
    \midrule
    0.6 & 0.570 & 1.922 & 670.2 & 2010.5 & $-616.3$ & 455.2 \\
    0.7 & 0.669 & 1.102 & 592.3 & 1776.8 & $-468.9$ & 368.8 \\
    0.8 & 0.769 & 0.675 & 536.3 & 1608.8 & $-365.0$ & 304.8 \\
    \bottomrule
  \end{tabular}
\end{table}

In the nuclear medium, we adopt the quark-meson coupling (QMC) model with the gaussian quark wavefunction, where the mean fields of $\sigma$, $\omega$, and $\rho$ mesons ($\bar{\sigma}$, $\bar{\omega}$, and $\bar{\rho}$) interact with the confined quarks, in uniformly distributed, asymmetric nuclear matter.
Under the relativistic HO potential given in Eq.~\eqref{eq:conpot}, the Dirac equation for the quark field $\psi_j$ ($j= u\, \textrm{or}\, d$) is given by~\cite{PhysRevD.110.113001}
\begin{equation}
  \left[i\gamma \cdot \partial -\left(m -V_{s}\right)-\gamma_{0}V_{0}-\frac{c}{2}\left( 1+\gamma_{0}\right) r^2 \right]\psi_{j}({\bm r}, t)=0,
  \label{eq:Dirac-eq1}
\end{equation}
where $V_{s} = g^{q}_{\sigma}\bar{\sigma}$ and $V_{0} = g^{q}_{\omega}\bar{\omega} \pm g^{q}_{\rho} \bar{\rho}$ for $\binom{u}{d}$ quark with the quark-meson coupling constants, $g^{q}_{\sigma}$, $g^{q}_{\omega}$ and $g^{q}_{\rho}$.
% Note that the isoscalar $\sigma$ meson couples to the $u$ and $d$ quarks equally.
We respectively define the effective quark mass and the effective single-particle quark energy as $m^{\ast} \equiv m-V_{s} = m - g^{q}_{\sigma}\bar{\sigma}$ and $\epsilon^{\ast} \equiv \epsilon_j -V_{0} = \epsilon_j - g^{q}_{\omega}\bar{\omega} \mp g^{q}_{\rho} \bar{\rho}$ for $\binom{u}{d}$ quark, where $\epsilon_{j}$ is the eigenenergy of Eq.~\eqref{eq:Dirac-eq1}.
The static, lowest-state wavefunction in matter is presented by $\psi_{j}({\bm r}, t) = \exp\left( -i\epsilon_{j}t \right) \psi_Q({\bm r})$.
The wavefunction $\psi_Q({\bm r})$ is then given by Eqs.~\eqref{eq:sol} and \eqref{eq:norm}, in which $\epsilon$, $m$, $\lambda$ and $a$ are, respectively, replaced with $\epsilon^{\ast}$, $m^{\ast}$, $\lambda^{\ast}$ and $a^{\ast}$, and the effective energy $\epsilon^{\ast}$ is determined by $\sqrt{\epsilon^\ast + m^\ast} (\epsilon^\ast - m^\ast) = 3 \sqrt{c}$.
The zeroth-order energy of the nucleon in matter is thus given by $E_N^{0 \ast} = 3 \epsilon^\ast$, and we have the effective nucleon mass in matter
\begin{equation}
  M_N^\ast = E_B^{0 \ast} + E^{spin}_N - E^{c.m. \ast}_N.
  \label{eq:nmass1}
\end{equation}
Here, we assume that the strength parameter, $c$, of the confinement potential and the spin correlation $E^{spin}_N$ do not change in matter.

\section{Quarkyonic picture}
\label{subsec:quarkyonic}

In symmetric nuclear matter (SNM), the dual momentum-space distribution for isospin-symmetric light quarks with a given color is expressed by the sum rule~\citep{PhysRevD.104.074005,PhysRevLett.132.112701}
\begin{equation}
  f_Q(q) = \int_k \varphi\left( {\bm q} - \frac{\bm k}{N_c} \right) f_N(k),
  \label{eq:sumrule}
\end{equation}
where $\int_k \equiv \int d^3{\bm k}/(2\pi)^3$, $f_N(k) \, [f_Q(q)]$ is the momentum-space distribution of nucleon $N$ [quark $Q$ ($=$ up ($u$) or down ($d$))] with Fermi statistics, $0 \leq f_N(k), f_Q(q) \leq 1$, and $N_c$ is the number of color.
% Hereafter, the letters $q$ and $k$ are used exclusively for quarks and nucleons, respectively.

\begin{figure}[t!]
  \centering
  \includegraphics[width=8.5cm,keepaspectratio,clip]{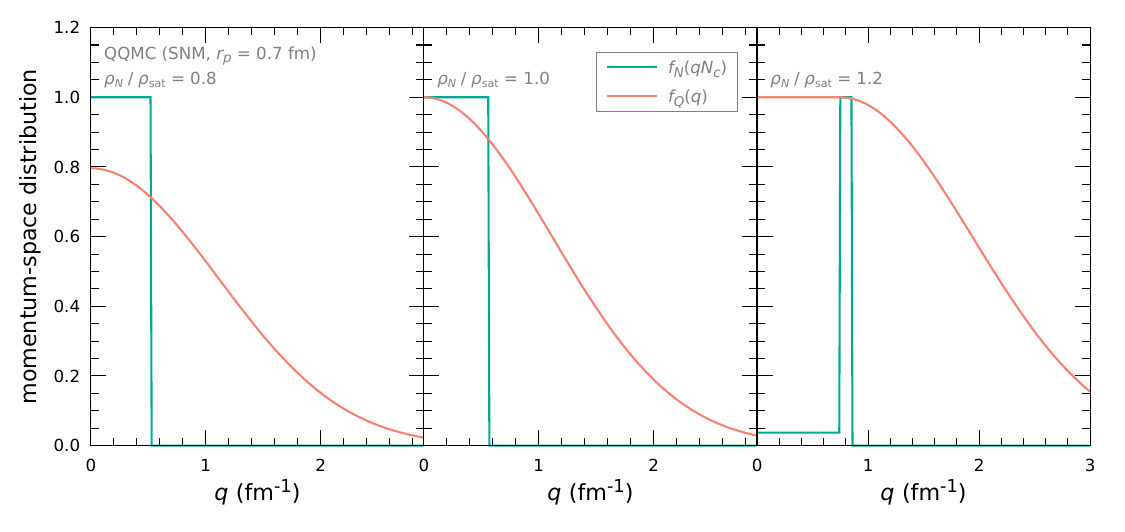}
  \caption{\label{fig:1}
    Quark and nucleon momentum-space distributions, $f_{Q}(q)$ and $f_{N}(k=qN_{c})$, in symmetric nuclear matter (SNM) for the case of $r_{p}=0.7$ fm using the quarkyonic QMC (QQMC) model.
    Above $\rho_{\rm sat}$, $f_{Q}(q)$ is supposed to be the {\it shifted} distribution to describe that the low momentum levels of quarks are fully occupied, as discussed in Ref.~\citep{PhysRevD.104.074005}.}
\end{figure}
At low density, the matter is best viewed in terms of only nucleons.
As shown in the left panel of Fig.~\ref{fig:1}, $f_N(k)$ is the usual Fermi distribution: $f_N(k) = \theta(k_s - k)$ with $k_s = k_F$ ($k_F$ the Fermi momentum).
Using the gaussian quark wavefunction in Eq.~\eqref{eq:sol}, the single quark momentum distribution in a single nucleon is given by
\begin{equation}
  \varphi({\bm q}) = \pi^{3/2} a^3 \left( \frac{16 \lambda^2 a^2}{2 \lambda^2 a^2 +3} \right)
  \left( 1+ \frac{q^2}{\lambda^2} \right) e^{- a^2 q^2}.
  \label{eq:qmom}
\end{equation}
Taking into account the sum rule in Eq.~\eqref{eq:sumrule}, the quark momentum-space distribution, $f_{Q}(q,k_{s})$, is calculated analytically and is less than unity~\citep{Saito:2025yld}.
Here, we explicitly present a variable  $k_{s}$ as well as $q$ to assign the nuclear density,
% $\rho_N = \frac{2}{\pi^2} \int dk\, k^2 f_N(k)=\frac{2k_{s}^{3}}{3\pi^{2}}$.
$\rho_N = \frac{2}{3\pi^{2}} k_{s}^{3}$.

As $\rho_N$ grows up, $f_Q(q,k_{s})$ increases and eventually reaches the upper bound.
In the middle panel of Fig.~\ref{fig:1}, the level of $q=0$ only is fully occupied, i.e. $f_Q(0,k_{s}) = 1$.
Above such density, the low momentum states of the quarks are fully occupied, i.e. {\em the quark saturation} occurs, and the matter undergoes a transition from ordinary nuclear matter to quarkyonic matter~\citep{MCLERRAN200783}.
We call this density the quark saturation density, $\rho_{\rm sat}$, and this phase may be regarded as {\em a soft deconfinement} region~\citep{PhysRevD.102.096017}.

As shown in the right panel of Fig.~\ref{fig:1}, above $\rho_{\rm sat}$, the dual nucleon momentum distribution in SNM is separated into two segments
\begin{equation}
  f_N(k) = \frac{1}{N_c^3} \theta(k_b - k) + \theta(k_s - k) \theta(k - k_b),
  \label{eq:postnucleon}
\end{equation}
which is the order of $1/N_c^3$ in the {\em bulk} Fermi sea ($k \leq k_b$) while forms the nucleon {\em shell} structure at $k_b \leq k \leq k_s$.
Substituting $f_{N}(k)$ into the sum rule in Eq.~\eqref{eq:sumrule}, the nucleon bulk and shell momenta, $k_{b}$ and $k_{s}$, can be calculated through the boundary condition: $f_{Q}(0,k_s)-\beta f_{Q}(0,k_b)=1$ with $\beta=(1-1/N_{c}^{3})$.
This boundary condition is the {\em necessary} condition, and is identical to the condition that sets the relation between $k_b$ and $k_s$ in the IdylliQ model~\cite{PhysRevLett.132.112701}.
Due to the Pauli blocking at the quark level, the shape of $f_{N}(k)$ in Eq.~\eqref{eq:postnucleon} is energetically more favorable than in the ordinary shell structure.
The $\rho_{N}$ is then modified as $\rho_{N}= \frac{2}{3\pi^{2}} (k_{s}^{3}-\beta k_{b}^{3})$.
This implies that the pressure and the sound velocity are considerably enhanced above $\rho_{\rm sat}$~\citep{PhysRevC.110.025201}.

In addition, we construct the quarkyonic QMC (QQMC) model, in which the nuclear interaction is involved at the level of mean-field approximation, replacing  $\epsilon$, $m$, $\lambda$ and $a$ in the gaussian wavefunction with the effective quantities, $\epsilon^{\ast}$, $m^{\ast}$, $\lambda^{\ast}$ and $a^{\ast}$~\citep{Saito:2025yld}.

\section{Numerical results}
\label{sec:results}

\begin{figure}[t!]
  \centering
  \includegraphics[width=8.0cm,keepaspectratio,clip]{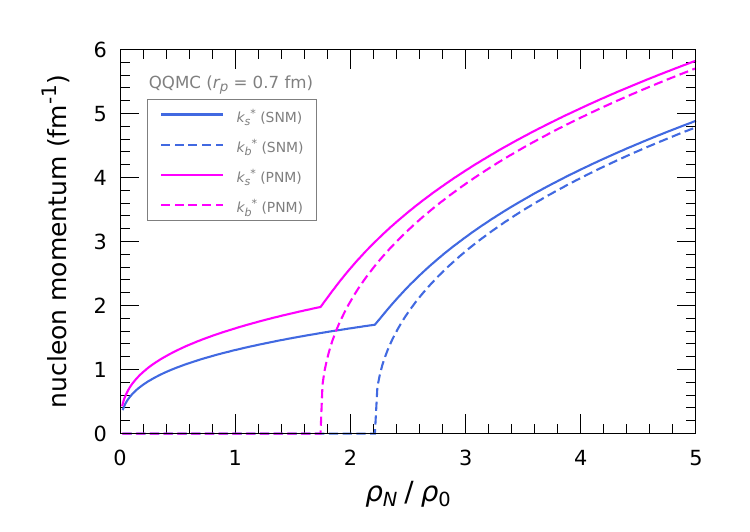}
  \caption{\label{fig:2}
    Density dependence of nucleon bulk and shell momenta, $k_{b}^{\ast}$ and $k_{s}^{\ast}$, in the quarkyonic QMC (QQMC) model.}
\end{figure}
In Fig.\,\ref{fig:2}, we illustrate the density dependence of nucleon bulk and shell momenta, $k_{b}^{\ast}$ and $k_{s}^{\ast}$, in symmetric nuclear matter (SNM) or in pure neutron matter (PNM) using the QQMC model.
The rapid rise of the momenta, $k_{b}^{\ast}$ and $k_{s}^{\ast}$, occurs beyond $\rho_{\rm sat}$, which leads to singular behavior in thermodynamic quantities.
With increasing $\rho_{N}$, the shell part shrinks while the bulk part dominates, as explained in the right panel of Fig.~\ref{fig:1}.
We find that in the quarkyonic phase, quarks with low momenta are suppressed due to the Pauli blocking at the quark level, which correspondingly leads to the suppression of slow nucleons and the increase of nucleons with relatively high momenta.

\begin{table}[b!]
  \centering
  \caption{\label{tab:CCs}
    Coupling constants and the quark saturation density, $\rho_{\rm sat}$, in the QQMC model.
    The nuclear saturation density, $\rho_{0}$, is chosen to be $\rho_{0}=0.15$ fm$^{-3}$.
    The $\rho_{\rm sat}$ in the GQ model is also listed in the last column.}
  \begin{tabular}{ccccccc}
    \toprule
    $r_p$ & $g_{\sigma}$ & $g_{\omega}$ & $g_{\rho}$ & $g_{2}$     & \multicolumn{2}{c}{$\rho_{\rm sat}/\rho_{0}$} \\
    \cline{6-7}
    (fm)  & \            &              &            & (fm$^{-1}$) & QQMC  & GQ   \\
    \midrule
    0.6   & 11.00        & 7.47         & 4.36       & 22.78       & 3.59  & 4.18 \\
    0.7   & 10.49        & 7.56         & 4.36       & 23.81       & 2.21  & 2.53 \\
    0.8   & 10.11        & 7.65         & 4.35       & 24.65       & 1.46  & 1.63 \\
    \bottomrule
  \end{tabular}
\end{table}
The total energy density in SNM is expressed as
\begin{equation}
  \varepsilon_{tot}
  = \varepsilon_{N}
  + \frac{1}{2}\left(m_{\sigma}^{2}\bar{\sigma}^{2}+m_{\omega}^{2}\bar{\omega}^{2}+m_{\rho\,}^{2}\bar{\rho}^{2}\right)
  + \frac{1}{3}g_{2}\bar{\sigma}^{3},
  \label{eq:energydensity}
\end{equation}
where
\begin{equation}
  \varepsilon_{N} = \frac{2}{\pi^{2}} \int dk \,k^{2} \sqrt{M_{N}^{\ast 2}+k^{2}} f_{N}(k),
  \label{eq:energykin}
\end{equation}
with the usual Fermi distribution for $\rho_{N} \le \rho_{\rm sat}$ and the dual nucleon momentum distribution in Eq.~\eqref{eq:postnucleon} for $\rho_{N} > \rho_{\rm sat}$.
The nucleon-meson coupling constants, $g_\sigma$,  $g_\omega$, and $g_\rho$, are respectively related to the quark-meson coupling constants as $g_\sigma = 3g_\sigma^q$,  $g_\omega = 3g_\omega^q$, and $g_\rho = g_\rho^q$.
The meson masses are taken to be $m_\sigma = 550$ MeV, $m_\omega = 783$ MeV, and $m_\rho = 770$ MeV.
The last term in Eq.\eqref{eq:energydensity} is the nonlinear, self-coupling term of $\sigma$ meson.
The coupling constants are listed in Table.~\ref{tab:CCs}.

\begin{figure}[t!]
  \centering
  \includegraphics[width=8.0cm,keepaspectratio,clip]{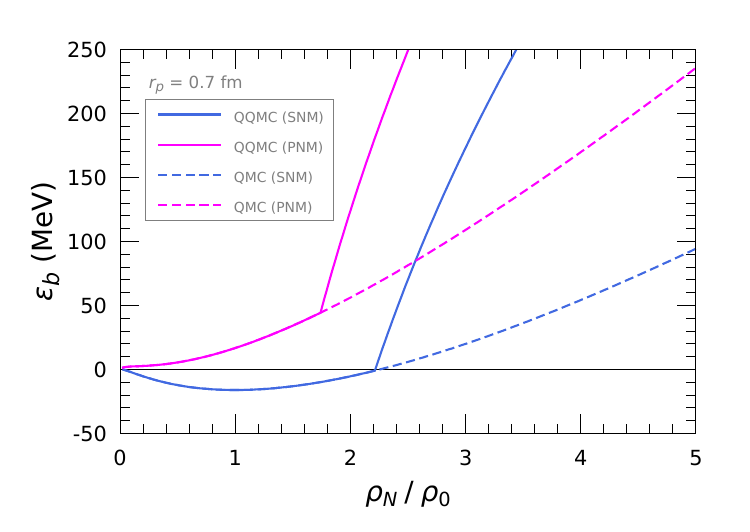}
  \caption{\label{fig:3}
    Binding energy per nucleon, $\varepsilon_{b}=\varepsilon_{\rm tot}/\rho_{N}-M_{N}$, as a function of baryon density ratio, $\rho_{N}/\rho_{0}$, in the QMC and QQMC models.}
\end{figure}
In Fig.~\ref{fig:3} we show the binding energy per nucleon, $\varepsilon_{b}=\varepsilon_{\rm tot}/\rho_{N}-M_{N}$, in the QMC and QQMC models for $r_p=0.7$ fm.
In the QQMC model, $\varepsilon_{b}$ rises rapidly once $\rho_{N}$ exceeds $\rho_{\rm sat}$, and this tendency is more pronounced in PNM than in SNM.
This is because, in PNM, the $d$-quark states are filled more rapidly than the $u$-quark states, so that the quark saturation sets in earlier.
For details, see Ref.~\citep{Saito:2025yld}.
As a consequence, the emergence of the quarkyonic phase strongly stiffens the nuclear EoS.

\begin{figure}[t!]
  \centering
  \includegraphics[width=7.8cm,keepaspectratio,clip]{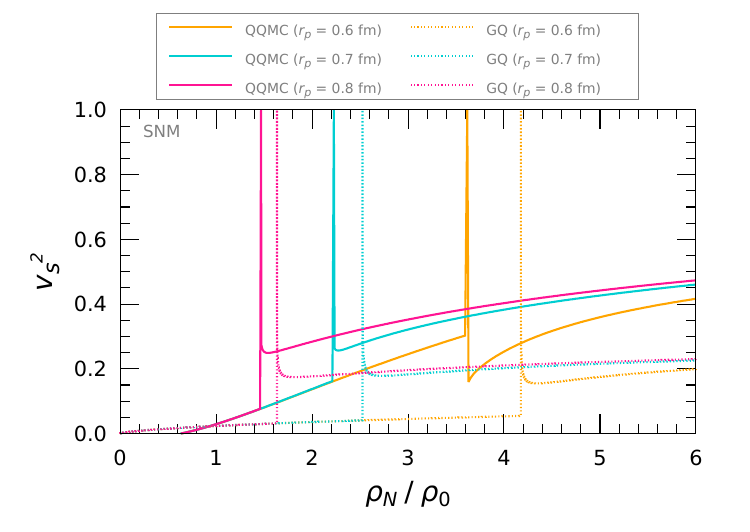}
  \caption{\label{fig:4}
    Sound velocity, $v_{s}^{2}$, as a function of baryon density ratio, $\rho_{N}/\rho_{0}$, for $r_p=0.6$, $0.7$, and $0.8$ fm in the QQMC and GQ models.}
\end{figure}
% Figure~\ref{fig:4} compares the sound velocity, $v_{s}^{2}$, in the QQMC model with that in the GQ model.
Figure~\ref{fig:4} shows the sound velocity, $v_{s}^{2}$, in the QQMC and GQ models.
In both models, $v_{s}^{2}$ exhibits singular behavior at $\rho_{\rm sat}$.
This divergence originates from the non-analytic structure of the momentum-space distribution and the strict duality constraint in the present idealized model.
It should not be interpreted as a physical phenomenon, but rather as an artifact of the sharp Fermi surface assumption.
The relativistic treatment shifts $\rho_{\rm sat}$ to higher density than in the nonrelativistic case~\citep{PhysRevC.110.025201}, whereas the inclusion of nuclear interactions slightly shifts $\rho_{\rm sat}$ back to lower density.
Nevertheless, $\rho_{\rm sat} > \rho_{0}$ holds in all cases considered here.

\section{Summary and conclusion}
\label{sec:summary}

In the present study, we have studied quarkyonic matter within a relativistic quark model by extending the QMC model to include interaction effects in dense matter.
The nucleon was described by relativistic gaussian quark wavefunctions, and the resulting QQMC model was applied to SNM and PNM.
We found that the quark saturation density strongly depends on the nucleon size parameter, and that the onset of quark saturation occurs earlier in the QQMC model than in the noninteracting GQ model.
This demonstrates that nuclear interactions play an important quantitative role in stiffening the nuclear EoS in the quarkyonic regime.
We have also shown that the increase in the binding energy and the singular behavior of the sound velocity are closely related to the emergence of quark saturation.
A future challenge is to avoid the divergence at the quark saturation density and to develop a smoother and more realistic description of quarkyonic matter~\citep{Saito:2025yld}.

\section*{Acknowledgments}

This work was supported by the Basic Science Research Program through the National Research Foundation of Korea (NRF) under Grant Nos. RS-2026-25471436, RS-2025-16071941, RS-2023-00242196, and RS-2021-NR060129.

\bibliographystyle{elsarticle-num-names-mod.bst}
\bibliography{Baryons2025.bib}

\end{document}